\documentclass[aps,prl,twocolumn,showpacs]{revtex4}
\usepackage{amssymb}
\usepackage{amsmath}
\usepackage{graphicx}
\usepackage{color}

\begin{document}

\title{Exact bosonization for an interacting Fermi gas in arbitrary
dimensions}
\author{K. B. Efetov${}^{1}$}
\author{C. P\'{e}pin${}^{2}$}
\author{H. Meier${}^{1}$}
\affiliation{$^{1}$Theoretische Physik III, Ruhr-Universit\"{a}t Bochum, 44780 Bochum,
Germany\\
$^{2}$IPhT, CEA-Saclay, L'Orme des Merisiers, 91191 Gif-sur-Yvette, France\\
}
\date{\today }

\begin{abstract}
We present an exact mapping of models of interacting fermions onto boson
models. The bosons correspond to collective excitations in the initial
fermionic models. This bosonization is applicable in any dimension and for
any interaction between fermions. Introducing superfields we derive a field
theory that may serve as a new way of analytical study. We show
schematically how the mapping can be used for Monte Carlo calculations and
argue that it should be free from the sign problem.
\end{abstract}

\maketitle

The study of interacting fermionic systems in cases when the Landau Fermi
liquid theory fails to describe all interesting effects is an open problem
of condensed matter theory. Very often conventional methods \cite{agd} are
not efficient due to divergencies in perturbation expansions leading to the
re-summation of complicated series.

It is difficult to list here all the problems encountered in the study of
e.g., strongly correlated systems. The most clear examples are provided by
one dimensional (1D) systems where perturbative methods are especially
inconvenient but models suggested for describing high $T_{c}$
superconductivity, see, e.g. \cite{lee}, are not simpler. Generically, the
low temperature physics of systems of interacting fermions is naturally
described in terms of bosonic collective excitations that can be expressed
only by infinite series of conventional diagrams.

The numerical study of fermionic systems encounters difficulties as well.
The powerful Monte Carlo (MC) method suffers the well known sign problem
\cite{blankenbecler,linden,hirsch,troyer,dosantos} leading to a drastic
increase in the computing time.

All these examples call for a reformulation of interacting fermions in terms
of a boson model. Such an approach, called bosonization (see, e.g. \cite%
{tsvelik,giamarchi}), is well known and successful for 1D fermionic systems.
Attempts to bosonize fermionic models in the dimensionality $d>1$ have been
undertaken in the past, starting from the works \cite{luther,haldane}
followed by \cite{houghton1}. \textbf{\ }These schemes, however, have
problems when dealing with large momentum transfer by the interaction \cite%
{castellani}. They reproduce the random phase approximation (RPA) but do not
reach beyond.

A more general low energy bosonization scheme suggested recently \cite%
{aleiner} is based on quasiclassical equations and takes into account large
momentum transfer. New logarithmic contributions to anomalous dependence of
the specific heat \cite{aleiner} and spin susceptibility \cite{schwiete}
were found. However, working well for $d=1$ the scheme of Ref. \cite{aleiner}
is not completely accurate for $d>1$ missing some effects of the Fermi
surface curvature \cite{chubukov}.

All the previous bosonization methods are not exact; hence they cannot be
used for accurate numerical studies of the initial fermionic problem.

In this paper, we present a new scheme that allows one to map interacting
fermions to interacting bosons \textit{exactly}. This mapping works in any
dimension at any temperature. The effective model obtained describes
interacting bosonic excitations. It can be written either in a form of a
model of non-interacting bosons in a Hubbard-Stratonovich (HS) field with a
subsequent integration over this field or in a form of a field theory
containing superfields with quartic and cubic interactions. The former may
be convenient for MC study, while the latter promises to be good for
analytical investigations.

We start with a general model of interacting electrons described by the
Hamiltonian
\begin{equation}
\hat{H}=\hat{H}_{0}+\hat{H}_{int},  \label{a1}
\end{equation}%
where $\hat{H}_{0}$ is the bare part,%
\begin{equation}
\hat{H}_{0}=-\sum_{r,r^{\prime },\sigma }t_{r,r^{\prime }}c_{r\sigma
}^{+}c_{r^{\prime }\sigma }-\mu \sum_{r,\sigma }c_{r\sigma }^{+}c_{r\sigma },
\label{a2}
\end{equation}%
and $\hat{H}_{int}$ stands for an electron-electron interaction,%
\begin{equation}
\hat{H}_{int}=\frac{1}{2}\sum_{r,r^{\prime }\sigma ,\sigma ^{\prime
}}V_{r,r^{\prime }}c_{r\sigma }^{+}c_{r^{\prime }\sigma ^{\prime
}}^{+}c_{r^{\prime }\sigma ^{\prime }}c_{r\sigma }.  \label{a3}
\end{equation}%
In Eqs. (\ref{a1}-\ref{a3}) $c_{r\sigma }$ $\left( c_{r\sigma }^{+}\right) $
are annihilation (creation) operator\textbf{s} of the electrons on a lattice
site $r$ with spin $\sigma =\pm $. The function $t_{r,r^{\prime }}$
describes the tunneling from the site $r$ to the site $r^{\prime },$ $%
V_{r,r^{\prime }}$ is the electron-electron interaction between the $r$ and $%
r^{\prime }$ and $\mu $ is the chemical potential.

The scheme of the bosonization suggested here can be developed for arbitrary
functions $t_{rr^{\prime }}$ and $V_{rr^{\prime }}$ in an arbitrary
dimension. However, in order to make formulas more compact we assume that
\begin{equation}
V_{r,r^{\prime }}=\delta _{r,r^{\prime }}V_{0},\quad V_0>0 ,  \label{a4}
\end{equation}%
which corresponds to an onsite repulsion.

Then, the term $\hat{H}_{int}$ can be rewritten in the form
\begin{equation}
\hat{H}_{int}^{\left( 0\right) }=-\frac{V_{0}}{2}\sum_{r}\left(
c_{r,+}^{+}c_{r,+}-c_{r,-}^{+}c_{r,-}\right) ^{2}  \label{a5}
\end{equation}%
while replacing the chemical potential $\mu$ by $\mu ^{\prime }= \mu -V_0/2.$

In this paper, we concentrate on studying thermodynamics and calculate the
partition function
\begin{equation}
Z=\mathrm{Tr}\exp \left( -\beta \hat{H}\right) ,\quad \beta =1/T.  \label{a6}
\end{equation}%
As $\hat{H}_{0}$ and $H_{int}^{\left( 0\right) }$ do not commute we
subdivide the interval $\left( 0,\beta \right) $ into slices of the length $%
\Delta =\beta /N\ll \beta $ and write $Z$ as a time ordered product over the
imaginary time $\tau $. Following a standard route of the HS transformation
we decouple the interaction term $\hat{H}_{int}^{\left( 0\right) }$
integrating over a periodic real field $\phi _{r}\left( \tau \right) =\phi
_{r}\left( \tau +\beta \right) $ and come to%
\begin{eqnarray}
Z &=&\lim_{\Delta \rightarrow 0}\int Z\left[ \phi \right] \exp \Big[-\frac{%
\Delta }{2V_{0}}\sum_{r}\sum_{l=1}^{N}\phi _{r,l}^{2}\Big]D\phi ,  \label{a7}
\\
Z\left[ \phi \right] &=&\mathrm{Tr}\Big[\mathrm{\exp }\Big(-\frac{\hat{H}_{0}%
}{T}\Big)\prod_{l=1}^{N}\exp \Big(\Delta \sum_{r,\sigma }\sigma \phi _{r,l}%
\bar{c}_{r,\sigma ,l}c_{r,\sigma ,l}\Big)\Big],  \notag
\end{eqnarray}%
$c_{r,\sigma ,l}$ and $\bar{c}_{r,\sigma ,l}$ are the annihilation and
creation operators in the interaction representation\cite{agd} taken at $%
\tau _{l}=\left( l-1/2\right) \Delta $, $D\phi $ is the normalized product
of all differentials $d\phi _{r,l}$ and $\phi _{r,l}=\phi _{r}\left( \tau
_{l}\right) $. The product in $Z\left[ \phi \right] $ is ordered in time,
such that $l=1$ is on the right.

For the analytical study, we could write Eq. (\ref{a7}) explicitly in the
continuous limit $\Delta \rightarrow 0$ using integrals and time ordering
operators $T_{\tau }$. However, MC calculations imply finite $\Delta $ with
typical values of $\phi _{r,l}$ growing as $\Delta ^{-1/2}$ for $\Delta
\rightarrow 0$ and therefore, we keep finite $\Delta $.

Calculation of the trace over the fermionic operators in Eq. (\ref{a7}) is
not simple for finite $\Delta $ and one should approximate $Z\left[ \phi %
\right] $ by a more convenient expression. A standard approximation $Z_{f}%
\left[ \phi \right] $ used in MC simulations instead of $Z\left[ \phi \right]
,$ Eq. (\ref{a7}), reads \cite{blankenbecler,dosantos}
\begin{eqnarray}
Z_{f}\left[ \phi \right] &=&\det_{r,\sigma }\Big[1+\prod_{l=1}^{N}\exp
\left( -\hat{h}\left[ \phi \left( \tau _{l}\right) \right] \Delta \right) %
\Big],  \label{b5} \\
\hat{h}_{r,\sigma }\left[ \phi \left( \tau \right) \right] &=&\hat{%
\varepsilon}_{r}-\mu ^{\prime }-\sigma \phi _{r}\left( \tau \right)  \notag
\end{eqnarray}%
where $\hat{\varepsilon}_{r}f_{r}\equiv -\sum_{r^{\prime }}t_{r,r^{\prime
}}f_{r^{\prime }}$ for an arbitrary function $f_{r}$.

We suggest here another approximation $Z_{b}\left[ \phi \right] $ to $Z\left[
\phi \right] $ that is more suitable for the bosonization,
\begin{eqnarray}
Z_{b}\left[ \phi \right]  &=&\mathrm{Tr}\Big[\mathrm{\exp }\Big(-\frac{\hat{H%
}_{0}}{T}\Big)  \label{b6} \\
&&\times T_{\tau }\exp \Big(\sum_{r,\sigma }\int_{0}^{\beta }\sigma \tilde{%
\phi}_{r}\left( \tau \right) \bar{c}_{r,\sigma }\left( \tau \right)
c_{r,\sigma }\left( \tau \right) d\tau \Big)\Big],  \notag \\
\tilde{\phi}_{r}\left( \tau \right)  &=&\phi _{r,l}\text{ \textrm{for} }%
\left( l-1\right) \Delta \leq \tau <l\Delta .  \notag
\end{eqnarray}%
The functional $Z_{b}\left[ \phi \right]$, Eq. (\ref{b6}), differs from $Z%
\left[ \phi \right] $, Eq. (\ref{a7}), by integration of the operator $\bar{c%
}_{r,\sigma }\left( \tau \right) c_{r,\sigma }\left( \tau \right) $ over
each slice instead of taking it in the middle of the slice and multiplying
by $\Delta $. Therefore, the difference between $Z_{b}\left[ \phi \right] $
and $Z\left[ \phi \right] $ should vanish in the limit $\Delta \rightarrow 0$%
. We emphasize, however, that $Z\left[ \phi \right] $, $Z_{f}\left[ \phi %
\right] $, and $Z_{b}\left[ \phi \right] $ differ from each other at finite $%
\Delta $. The functional $Z_{b}\left[ \phi \right] ,$ Eq. (\ref{b6}), has a
form of the exact partition function for an electron in an external
(generally, discontinuous in time) field $\tilde{\phi}_{r}\left( \tau
\right) $ and we can use standard transformations.

In order to reduce the fermionic model, Eqs. (\ref{b6}), to a bosonic one,
we introduce as in Ref. \cite{aleiner} an additional variable $0\leq u\leq 1$
and write the function $Z_{b}\left[ \phi \right] $ as
\begin{equation*}
Z_{b}\left[ \phi \right] =Z_{0}\exp \Big[\sum_{r,\sigma }\int_{0}^{\beta
}\int_{0}^{1}\sigma \tilde{\phi}_{r}\left( \tau \right) G_{r,r;\sigma
}^{\left( u\phi \right) }\left( \tau ,\tau +0\right) dud\tau \Big]
\end{equation*}%
where $Z_{0}$ is the partition function of the ideal Fermi gas and $%
G_{r,r^{\prime };\sigma }^{\left( u\phi \right) }\left( \tau ,\tau ^{\prime
}\right) $ is a fermionic Green function,

\begin{equation}
\left( -\frac{\partial }{\partial \tau }-\hat{h}_{r,\sigma }\left[ u\tilde{%
\phi}\left( \tau \right) \right] \right) G_{r,r^{\prime };\sigma }^{\left(
u\phi \right) }\left( \tau ,\tau ^{\prime }\right) =\delta _{r,r^{\prime
}}\delta \left( \tau -\tau ^{\prime }\right) ,  \label{a10}
\end{equation}%
with the boundary conditions%
\begin{equation*}
G_{r,r^{\prime };\sigma }^{\left( u\phi \right) }\left( \tau ,\tau ^{\prime
}\right) =-G_{r,r^{\prime };\sigma }^{\left( u\phi \right) }\left( \tau
+\beta ,\tau ^{\prime }\right) =-G_{r,r^{\prime };\sigma }^{\left( u\phi
\right) }\left( \tau ,\tau ^{\prime }+\beta \right) .
\end{equation*}%
We develop our bosonization scheme introducing
\begin{equation}
A_{r,r^{\prime }}(z)=G_{r,r^{\prime }}^{\left( 0\right) }\left( \tau ,\tau
+0\right) -G_{r,r^{\prime };\sigma }^{\left( u\phi \right) }\left( \tau
,\tau +0\right) \ ,  \label{adef}
\end{equation}%
where $z=\left( \tau ,\sigma ,u\right) $ and $G_{r,r^{\prime }}^{\left(
0\right) }\left( \tau ,\tau ^{\prime }\right) $ is the bare electron Green
function. The function $A_{r,r^{\prime }}\left( \tau \right) $ is periodic, $%
A_{r,r^{\prime }}\left( \tau \right) =A_{r,r^{\prime }}\left( \tau +\beta
\right) $ , and, hence, describes bosons.

We rewrite the partition function $Z\left[ \phi \right] $ as
\begin{equation}
Z_{b}\left[ \phi \right] =Z_{0}\exp \Big[-\sum_{r,\sigma }\int_{0}^{\beta
}\int_{0}^{1}\sigma \tilde{\phi}_{r}\left( \tau \right) A_{r,r}\left(
z\right) dud\tau \Big].  \label{a18}
\end{equation}%
and derive a closed equation for $A_{r,r^{\prime }}\left( z\right) $. For
that purpose we write a conjugated equation
\begin{equation*}
\left( \frac{\partial }{\partial \tau ^{\prime }}-\hat{h}_{r^{\prime
},\sigma }\left[ u\tilde{\phi}\left( \tau ^{\prime }\right) \right] \right)
G_{r,r^{\prime };\sigma }^{\left( u\phi \right) }\left( \tau ,\tau ^{\prime
}\right) =\delta _{r,r^{\prime }}\delta \left( \tau -\tau ^{\prime }\right)
\end{equation*}%
and subtract it from Eq. (\ref{a10}) putting in the resulting equation $\tau
^{\prime }=\tau +0$. The same can be done for the bare Green function $%
G_{r,r^{\prime };\sigma }^{\left( 0\right) }\left( \tau ,\tau ^{\prime
}\right) $ and finally we obtain \textbf{\ }
\begin{eqnarray}
\frac{\partial }{\partial \tau }A_{r,r^{\prime }}\left( z\right) &+&\mathcal{%
H}_{r,r^{\prime }}\left( z\right) A_{r,r^{\prime }}\left( z\right) =-u\sigma
\tilde{\Phi}_{r,r^{\prime }}\left( \tau \right) n_{r,r^{\prime }},  \notag \\
\mathcal{H}_{r,r^{\prime }}\left( z\right) &=&\hat{\varepsilon}_{r}-\hat{%
\varepsilon}_{r^{\prime }}-u\sigma \tilde{\Phi}_{r,r^{\prime }}\left( \tau
\right) ,\text{ }  \label{a19a} \\
\tilde{\Phi}_{r,r^{\prime }}\left( \tau \right) &=&\tilde{\phi}_{r}\left(
\tau \right) -\tilde{\phi}_{r^{\prime }}\left( \tau \right)  \notag
\end{eqnarray}
The function $n_{r,r^{\prime }}=G_{r,r^{\prime }}^{\left( 0\right) }\left(
\tau ,\tau +0\right) $ in Eq. (\ref{a19a}) is the Fourier transform in $%
r-r^{\prime }$ of the Fermi distribution. Equation (\ref{a19a}) should be
supplemented by {the condition }%
\begin{equation}
\sum_{r}{A}_{r,r}{\left( z\right) =0.}  \label{b3}
\end{equation}%
{Equation (\ref{b3}) can be obtained noticing that }$\sum_{r}A_{r,r}\left(
z\right) $ is a constant {independent of }$\phi _{r}\left( \tau \right) ,$
which{\ follows from Eq.~(\ref{a19a})}. {Assuming that the interaction and,
hence, }$\phi _{r}\left( \tau \right) ${\ vanishes at infinity }we come to
Eq. (\ref{b3}){. }

So, we are to solve the linear equation (\ref{a19a}) for $A_{r,r^{\prime
}}\left( z\right) $ with the condition (\ref{b3}), substitute the solution
into Eq. (\ref{a18}) and then into the first equation (\ref{a7}). A possible
strong discontinuity of the function $\tilde{\phi}$ does not lead to any
problems in the limit $\Delta \rightarrow 0.$ Our scheme is similar to that
of Ref. \cite{aleiner} developed in the quasiclassical approximation but now
all the transformations are exact.

It is convenient to exactly integrate over the field $\phi _{r}\left( \tau
\right) $ in the beginning and thus derive a field theory for interacting
bosons. In Ref. \cite{aleiner} this goal has been achieved by integrating
over $48$-component supervectors, which has led to a rather cumbersome
Lagrangian. Now we use another trick, known as the Becchi-Rouet-Stora-Tuytin
(BRST) transformation, based on introducing superfields \cite{brst} (see
also the book \cite{justin}). A similar transformation was used in the
quantization of non-abelian gauge theories \cite{faddeev}. In condensed
matter physics this trick has been used first in Ref. \cite{parisi}.

Within this method one replaces solving an equation
\begin{equation}
F\left( A\right) =0,  \label{a25}
\end{equation}%
where $F$ is a matrix function of a matrix function $A,$ and a subsequent
calculation of a quantity $B(A_{0}),$ where $A_{0}$ is the solution of Eq. (%
\ref{a25}), by an integral of the form%
\begin{equation}
B=\int B(a)\ \delta \left[ F\left( a\right) \right] \left\vert \det \left(
\frac{\partial F}{\partial a}\right) \right\vert da.  \label{a26}
\end{equation}%
The $\delta $-function can be written as%
\begin{equation*}
\delta \left[ F\left( a\right) \right] =C\int \exp \left[ ifF\left( a\right) %
\right] df,
\end{equation*}%
where $C$ is a coefficient, and the determinant is obtained after
integration of an exponential of a quadratic form in Grassmann variables $%
\eta $ and $\eta^{+}$.

Our problem of solving Eq. (\ref{a19a}) and calculation of the integral in
Eq. (\ref{a18}) is of this type and we proceed following the above trick. We
introduce anticommuting variables $\theta $ and $\theta ^{\ast }$ and a
superfield $\Psi _{r,r^{\prime }}\left( R\right) $, $R=\left\{ z,\theta
,\theta ^{\ast }\right\} $,
\begin{equation*}
\Psi _{r,r^{\prime }}\left( R\right) =a_{r,r^{\prime }}\left( z\right)
\theta +f_{r,r^{\prime }}^{T}\left( z\right) \theta ^{\ast }+\eta
_{r,r^{\prime }}\left( z\right) +\eta _{r,r^{\prime }}^{+}\left( z\right)
\theta ^{\ast }\theta
\end{equation*}%
where $a,f$ are real and $\eta $ is an anticommuting field. The field $\Psi $
is periodic as a function of $\tau $, $\Psi \left( \tau \right) =\Psi \left(
\tau +\beta \right) $, but is anticommuting. The hermitian conjugation $``+"$
implies both the complex conjugation and transposition $``T"$ with respect
to $r,r^{\prime }$.

As a result, one comes to an effective action quadratic in $\Psi $ and
linear in $\phi _{r}(\tau )$. This allows us to integrate over $\phi
_{r}(\tau )$ with the Gaussian weight of Eq.~(\ref{a7}) and we come to the
final expression for the partition function $Z$,%
\begin{equation}
Z=Z_{0}\int \exp \left[ -S_{0}\left[ \Psi \right] -S_{B}\left[ \Psi \right]
-S_{I}\left[ \Psi \right] \right] D\Psi ,  \label{a29}
\end{equation}%
where $S_{0}\left[ \Psi \right] $ is the bare part of the action,%
\begin{equation*}
S_{0}=\frac{i}{2}\sum_{r,r^{\prime }}\int \Big[\Psi _{r^{\prime },r}\left(
\frac{\partial }{\partial \tau }+\left( \hat{\varepsilon}_{r}-\hat{%
\varepsilon}_{r^{\prime }}\right) \right) \Psi _{r,r^{\prime }}\Big]dR,
\end{equation*}%
and the interaction terms are given by
\begin{align*}
S_{B}& =-\frac{V_{0}}{2}\sum_{r}\int \delta (\tau -\tau _{1})\Psi
_{r,r}\left( R\right) \theta ^{\ast } \\
\times & \Big[\Psi _{r,r}\left( R_{1}\right) \theta _{1}^{\ast }+2i\Pi
_{r}\left( R_{1}\right) \Big]\sigma \sigma _{1}dRdR_{1}\,,
\end{align*}%
\begin{equation*}
S_{I}=\frac{V_{0}}{2}\sum_{r}\int \delta (\tau -\tau _{1})\Pi _{r}\left(
R\right) \Pi _{r}\left( R_{1}\right) \sigma \sigma _{1}dRdR_{1},
\end{equation*}%
\begin{equation*}
\Pi _{r}\left( R\right) =u\sum_{r^{\prime }}\Big[\left( \Psi _{r^{\prime
},r}\left( R\right) -n_{r^{\prime },r}\theta \right) \left( \Psi
_{r,r^{\prime }}\left( R\right) -n_{r,r^{\prime }}\theta \right) \Big]
\end{equation*}%
Integration over $R$ in Eq. (\ref{a29}) implies summation over $\sigma $ and
integration over $u,\tau ,\theta ,\theta ^{\ast }$. The bare action $S_{0}$
and the interaction term~$S_{I}$ are invariant under the transformation of
the fields $\Psi $
\begin{equation}
\Psi _{r,r^{\prime }}\left( \theta ,\theta ^{\ast }\right) \rightarrow \Psi
_{r,r^{\prime }}\left( \theta +\kappa ,\theta ^{\ast }+\kappa ^{\ast
}\right) -\kappa n_{r,r^{\prime }}  \label{aSUSY}
\end{equation}%
with $\kappa $ and $\kappa ^{\ast }$ being anticommuting variables, whereas
the term $S_{B}$ breaks the invariance. The invariance under the
transformation (\ref{aSUSY}) is stronger than the standard BRST symmetry for
stochastic field equations (invariance under the transformation $\Psi \left(
\theta ^{\ast }\right) \rightarrow \Psi \left( \theta ^{\ast }+\kappa ^{\ast
}\right) $), Ref.~\cite{justin}, and reflects additional symmetries of Eq. (%
\ref{a19a}). It differs from the full supersymmetry by the presence of the
term $\kappa n_{r,r^{\prime }}$ in Eq. (\ref{aSUSY}) but still can lead to
interesting Ward identities.

The model described by Eqs. (\ref{a29}) can be studied using standard
methods of field theory. One can, e.g., expand in the interaction $V_{0}$ or
develop a renormalization group scheme analogous to that of Ref. \cite%
{aleiner}. In both the cases one can use the Wick theorem with rather simple
contraction rules that follow from the form of the bare action $S_{0}$. We
leave such calculations for future publications.

Neglecting cubic and quartic in $\Psi $ terms in $S_{B}$ and $S_{I}$ in
Eqs.~(\ref{a29}) one has a purely quadratic action and the partition
function~$Z$ yields an RPA-like expression,
\begin{eqnarray}
&&Z\simeq Z_{0}\exp \left[ -\frac{T}{2}\sum_{\omega }\int \frac{d^{d}\mathbf{%
k}}{\left( 2\pi \right) ^{d}}\ln K\right] ,  \label{aRPA} \\
&&K=1+V_{0}\int \frac{n\left( \mathbf{p-k/}2\right) -n\left( \mathbf{p+k/}%
2\right) }{i\omega +\varepsilon \left( \mathbf{p-k}/2\right) -\varepsilon
\left( \mathbf{p+k}/2\right) }\frac{d^{d}\mathbf{p}}{\left( 2\pi \right) ^{d}%
}.  \notag
\end{eqnarray}%
The same result can be obtained using Eqs. (\ref{a18}, \ref{a19a}) and
neglecting the field~$\tilde{\phi}_{r}(\tau )$ in the L.H.S. of Eq.~(\ref%
{a19a}).

In Eq. (\ref{aRPA}), $\left( K-1\right) $ is the contribution of
non-interacting bosonic excitations. Considering their interaction
originating from the cubic and quartic in $\Psi $ term in Eqs. (\ref{a29})
one can fully describe the initial fermionic system. So, going beyond RPA,
Eq. (\ref{aRPA}), is straightforward and this is a very important advantage
with respect to the older bosonization schemes \cite%
{luther,haldane,houghton1,castellani}. We are confident that the present
scheme can improve the analysis of Ref. \cite{aleiner} of non-analytical
corrections to the Landau Fermi liquid theory and expect its usefulness for
study of a large variety of problems of strongly correlated systems.

Now we sketch a possible route for MC simulations. Standard MC algorithms
are based on using Eq. (\ref{b5}). However, for some important
configurations of $\phi _{r}\left( \tau \right) $ the fermionic determinant $%
Z_{f}\left[ \phi \left( \tau \right) \right] $ is negative, which makes the
MC procedure inefficient. This is the famous sign problem.

Here we suggest to use instead of $Z_{f}\left[ \phi \left( \tau \right) %
\right] $ the functional $Z_{b}\left[ \phi \left( \tau \right) \right] $,
Eq. (\ref{a18}), that can be found solving Eq. (\ref{a19a}). The solution of
Eq. (\ref{a19a}) and the function $Z_{b}\left[ \phi \left( \tau \right) %
\right] $ can be approximated using a Green function $\mathcal{G}%
_{r,r^{\prime };r_{1},r_{1}^{\prime }}^{\sigma ,u\phi }\left( \tau ,\tau
_{1}\right) $ introduced as the solution of equation
\begin{equation*}
\left( \frac{\partial }{\partial \tau }+\mathcal{H}_{r,r^{\prime }}\left(
\tau \right) \right) \ \mathcal{G}_{r,r^{\prime };r_{1},r_{1}^{\prime
}}^{\sigma ,u\phi }\left( \tau ,\tau _{1}\right) =\delta (\tau -\tau
_{1})\delta _{r,r_{1}}\delta _{r^{\prime },r_{1}^{\prime }}.
\end{equation*}%
Then, we write the functional $Z_{b}\left[ \phi \left( \tau \right) \right] $
as%
\begin{eqnarray}
Z_{b}\left[ \phi \left( \tau \right) \right]  &=&Z_{0}\exp \Big[%
-\sum_{\sigma ,r,r_{1},r_{1}^{\prime }}\sum_{i,j}\int_{0}^{1}\mathcal{G}%
_{r,r;r_{1},r_{1}^{\prime }}^{\sigma ,u\phi }\left( \tau _{i},\tau
_{j}\right)   \notag \\
&&\times \tilde{\phi}_{r}\left( \tau _{i}\right) n_{r_{1},r_{1}^{\prime }}%
\tilde{\Phi}_{r_{1},r_{1}^{\prime }}\left( \tau _{j}\right) \Delta ^{2}du%
\Big]\mathbf{.}  \label{a20}
\end{eqnarray}

Similarly to Eq. (\ref{b5}), we write the function $\mathcal{G}%
_{r,r;r_{1},r_{1}^{\prime }}^{\sigma ,u\phi }\left( \tau _{i},\tau
_{j}\right) $ for $\beta >\tau _{i}>\tau _{j}>0$ in the form%
\begin{eqnarray}
&&\mathcal{G}_{r,r^{\prime };r_{1},r_{1}^{\prime }}^{\sigma ,u\phi }\left(
\tau _{i},\tau _{j}\right) =\hat{P}_{r,r^{\prime }}\left( \tau _{i},\tau
_{j}\right)  \label{a21} \\
&&\times \left( 1-\hat{P}_{r,r^{\prime }}\left( \tau _{j},0\right) \hat{P}%
_{r,r^{\prime }}\left( \beta ,\tau _{j}\right) \right) ^{-1}\delta
_{r,r_{1}}\delta _{r^{\prime },r_{1}^{\prime }}.  \notag
\end{eqnarray}%
Herein, the operator $\hat{P}$ is given by the expression%
\begin{equation}
\hat{P}_{r,r^{\prime }}\left( \tau _{i},\tau _{j}\right) =\prod_{i\geq l\geq
j}\exp \left( -\mathcal{H}_{r,r^{\prime }}\left( \tau _{l},\sigma ,u\right)
\Delta \right)  \label{a22}
\end{equation}%
where the multipliers in the product are ordered in time growing from the
right to the left. (Of course, one should discretize also the variable $u$).
The function $\mathcal{G}_{r,r^{\prime };r_{1},r_{1}^{\prime }}^{\sigma
,u\phi }\left( \tau _{i},\tau _{j}\right) $ satisfies the symmetry relation%
\begin{equation}
\mathcal{G}_{r,r^{\prime };r_{1},r_{1}^{\prime }}^{\sigma ,u\phi }\left(
\tau _{i},\tau _{j}\right) =-\mathcal{G}_{r_{1}^{\prime },r_{1};r^{\prime
},r}^{\sigma ,u\phi }\left( \tau _{j},\tau _{i}\right) ,  \label{a23}
\end{equation}%
which allows one to consider times $\beta >\tau _{j}>\tau _{i}>0$.

The form of the Green function Eq.~(\ref{a21}) is typical for bosons. By
construction (see Eqs. (\ref{a19a}, \ref{a21})) it is real unless a
singularity is present, in which case an imaginary part might be generated.
We argue that a possible zero in the Bose-denominator in Eq.~(\ref{a21}) is
compensated by the function $\Phi _{r_{1},r_{1}^{\prime }}$, Eq.~(\ref{a19a}%
), vanishing at $r_{1}=r_{1}^{\prime }$. Alternatively, one can
antisymmetrize in the beginning the function $\mathcal{G}$ in $%
r_{1},r_{1}^{\prime }$ by antisymmetrizing the $\delta $-functions in Eq.~(%
\ref{a19a}). This compensation is clearly seen in the RPA, Eq.~(\ref{aRPA}).

In the absence of any singularity, the result is insensitive to the way of
subdividing the interval~$(0,\beta )$ into slices and $Z_{b}[\phi \left(
\tau \right) ]$ remains positive in the process of the calculation for any $%
\phi _{r}\left( \tau _{i}\right) $ excluding the sign problem. Since $Z$ can
now be expanded in a sum of positive terms, we believe that this MC
procedure can be efficient. The above derivation can be done using the
\textquotedblleft Ising spin" auxiliary field of Refs. \cite{hirsch,dosantos}
as well, which is usually preferable for MC computations.

In conclusion, the exact bosonization method presented here opens new
possibilities of both numerical and analytical study of models of
interacting fermions. There is a reasonable chance that this new formalism
is free from the sign problem supposed to be generically NP-hard \cite%
{troyer} or problems of equivalent complexity.

We thank Transregio 12 of DFG, and the French ANR for financial support. We
are grateful to F. David, A. Ferraz, O. Parcollet for very useful
discussions.

\end{document}